\documentclass{elsart}
\usepackage{amssymb}
\usepackage{amsmath}
\usepackage{graphics}

\usepackage{epsfig}
\usepackage{amssymb}

\begin{document}

\begin{frontmatter}
%\title{Deuteron breakup reaction $pd\to(pp)n$ at high transferred momentum
%and short-range NN properties}
\title{The reaction $pd\to(pp)n$ at high momentum transfer and short-range 
$NN$ properties}

%\title{The reaction $pd\to(pp)n$ and the short-range properties
%of the deuteron and diproton wave functions}

\author[ikp]{J.~Haidenbauer},
\author[dubna]{Yu.N.~Uzikov}
\footnote{E-mail address: uzikov@nusun.jinr.ru\\
Permanent address: Kazakh National University, 480078 Almaty, Kazakstan}

\vskip 5mm
\address[ikp]{Institut f\"ur Kernphysik, Forschungszentrum J\"ulich, 52425
 J\"ulich, Germany}
\address[dubna]{Joint Institute for Nuclear Research, LNP, 141980 Dubna,
 Moscow Region, Russia}

%\vspace{1cm}
%\newpage

\begin{abstract}
{
A recent cross section measurement of the deuteron breakup reaction
$pd\to (pp)(0^\circ)+n(180^\circ)$, in the GeV region in a kinematics 
similar to backward $pd$ elastic scattering, 
strongly contradicts predictions of a $pd \to dp$ model 
based on the one-nucleon exchange, single pN scattering and
$\Delta$ excitation mechanisms, and on the wave functions of
the Reid soft core and Paris $NN$ potentials.
We show within the same model
that for the CD Bonn $NN$ potential there is qualitative agreement 
with the data. It is attributed to 
a reduction of the one-nucleon exchange at energies above 1 GeV
and an increase of the $\Delta(1232)$-isobar contribution,
both related to the short-range properties of the
wave functions generated by this potential.
 }
\vspace{1cm}
\end{abstract}

%%%%%%%%%%%%%%%%%%%%%%%%%%%%%%%%%%%%%
\begin{keyword}
Deuteron breakup;  Short--range nucleon--nucleon interaction

\begin{PACS}
13.75.Cs, 25.10.+s, 25.40-h.\\[1ex]
\end{PACS}
\end{keyword}

%\begin{keyword}
%Elastic scattering;  Short-range interaction of nucleons

%\begin{PACS}
%13.75.Cs, 25.45.D, 25.10.+s\\[1ex]
%\end{PACS}
%\end{keyword}
\end{frontmatter}
\baselineskip 4ex
\newpage

 The  structure of the lightest nuclei at short distances in the nucleon
 overlap region $r_{NN}< 0.5 $ fm, {\it i. e.} at high relative momenta
 $q_{NN}\sim 1/r_{NN}> 0.4 $ GeV/c between the nucleons,
 can be tested by electromagnetic probes at high momentum transfer
\cite{garsonvanorden}.
 However, a self-consistent picture 
 of electro- and photo-nuclear processes is not yet developed, 
 mainly because of 
 the unknown strength of the meson-exchange currents (MEC). 
 Hadron-nucleus collisions can give, in principle, independent information.
 On the other hand, here the theoretical analysis is obstructed 
 by initial and final state interactions and by the excitation/de-excitation
 of nucleons in the intermediate states. 
 For example, a large contribution
 of double $pN$ scattering with excitation of the $\Delta(1232)$ 
 resonance was found in proton-deuteron 
 backward elastic scattering ($pd\to dp$) at
 beam energies $T_p \sim 0.5$ GeV
 \cite{cwilkin69,kls,bdillig,imuz88,umuzsc89,uz98}.
 At higher  energies it is expected that also heavier baryon resonances 
 will play an important role. Like  
 the MEC problem in electro-nuclear interactions,
 the nucleon-isobar contributions are theoretically not well 
 under control, due to the rather poor information about
 the $pN\rightleftharpoons NN^*$ and $pN\rightleftharpoons N\Delta$ amplitudes.
 These difficulties are the main reasons why expectations to consider
 the reaction $pd\to dp$ as a probe for the short-range structure of
 the deuteron could not be realized yet \cite{garsonvanorden}.

In order
 to minimize those complicating effects it was proposed in Ref. \cite{imuz90}
 to study the deuteron breakup reaction $pd\to (pp)n$ in a kinematics 
 similar to 
 backward elastic $pd$ scattering. For small excitation energies, 
 $E_{pp} \leq 3$ MeV, the final $pp$ pair (diproton) is expected to be mainly 
in the
$^1S_0$ spin singlet (isotriplet) state \cite{imuz90,smuz98}. In contrast to
 $pd\to dp$, the isovector nature of the diproton
 causes a considerable suppression of the
 $\Delta-$ (and $N^*$) excitation amplitude 
 in comparison with the one-nucleon exchange (ONE) contribution,
 due to the additional isospin factor $\frac{1}{3}$.
 The same suppression factor acts for a broad class
 of diagrams with isovector meson--nucleon rescattering in the intermediate
 state including the excitation of any baryon resonances \cite{uzzhetf}.
 Furthermore, the node in the
 $pp$ $(^1S_0)$ half-off-shell reaction amplitude
 at the off-shell momentum $q\approx 0.4$ GeV/c induces some remarkable
 features in spin observables and leads to a dip in the unpolarized
 cross section for the ONE mechanism \cite{imuz90,uz2002}. Though a similar
 node occurs also in the deuteron $S$-wave wave function, its influence 
 in the $pd\to dp$ and $pd\to pX$ processes is, however, hidden by the 
 large contribution of the $D$-wave component. 
 Thus, the specific features of the $pd\to (pp)n$ reaction mentioned above
 provide a new testing ground for 
the $pd$ dynamics at high momentum transfer 
 and, accordingly, for the properties of the commonly used $NN$ potentials  
 at short distances.

\begin{figure}[hbt]
\mbox{\epsfig{figure=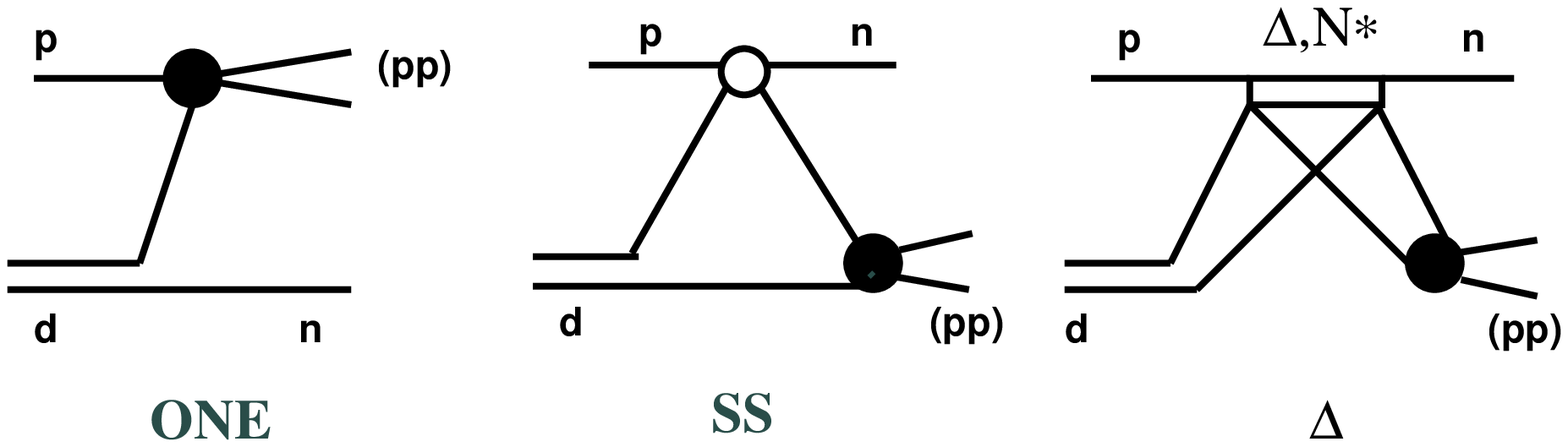,height=0.15\textheight, clip=}}
\caption{Mechanisms of the breakup reaction $pd\to (pp)n$.
The same mechanisms are used for the reaction $pd\to dp$.
}
\label{mechanism}
\end{figure}

 In the present paper we analyze 
 the first data on the reaction $pd\to (pp)n$, obtained 
 recently at ANKE-COSY \cite{vikomarov} for beam energies
 $T_p=0.6 - 1.9$ GeV with forward emission
 of a fast proton pair of low excitation energy of $E_{pp}=0-3$~MeV. 
 Existing predictions, produced in the framework of a model 
 which includes the triangle diagram of $pN$ single scattering (SS) in 
 addition to the ONE and $\Delta$-mechanisms
 (ONE+SS+$\Delta$, cf. Fig. \ref{mechanism}),
 and employing the RSC \cite{rsc} and Paris \cite{lacombe}
 $NN$ potentials \cite{imuz90}, 
 turned out to be in agreement with the now measured
cross section \cite{vikomarov} at low energy 0.6-0.7 GeV only.
 Specifically, the dip of the cross section at around $T_p \approx 0.8$ GeV,
 suggested by that model calculation \cite{imuz90,uz2002},  
 was not observed in the experiment. Moreover, at higher energies,
 $T_p>1 $~GeV, the predicted cross section exceeds the data by 
 a factor of 2--3. 
 According to an analysis presented in Ref. \cite{ukrs02}
 those two deficiencies could be a consequence of a too large contribution
 from the ONE mechanism.
 Distortions reduce the ONE contribution and, in turn, improve the
 results somewhat \cite{ukrs02},
 but they do not really remove the disagreement of the model calculation
 with the data.
 Therefore, it was argued in Ref. \cite{ukrs02} that the high-momentum
 components of the $NN$ wave functions should be much smaller as those
 of the employed RSC and Paris potentials in order to achieve agreement with the
 experiment, or in other words those wave functions should be soft.
 In the present work we show, within the ONE+SS+$\Delta$ model, that with the
 use of
 wave functions generated from the CD Bonn $NN$ potential \cite{machl}
 a qualitative agreement between the 
 calculations
 and the breakup data can be obtained,   
 especially after taking into account initial and final state
 interactions for the ONE mechanism.

 The cms 3-fold differential cross section of the reaction $pd\to (pp)n$
 is given by \cite{smuz98}
\begin{equation}
\label{3fold}
\frac{d^3\sigma}{dk^2 d\Omega_n}=\frac{1}{(4\pi)^5} \frac{p_n}{p_i}
\frac{k}{s\, \sqrt{m^2+k^2}}\frac{1}{2}\int\int d\Omega_{\bf k}
{\overline {|M_{fi}|^2}} \ .
\end{equation}
% and  In Eq. (\ref{3fold})
 Here $p_i$ and $p_n$ are the cms momenta of the incident proton and the
 final neutron, respectively,
 $s$ is the squared invariant mass of the $p+d$ system, and ${\bf k}$ is the relative
 momentum in the final $pp$ system. The latter is related
 to the relative energy 
 in the $pp$ system, $E_{pp}$, by $k^2=m\,E_{pp}$ where $m$ is the
 nucleon mass. 
$M_{fi}$ is the matrix element of the reaction. 
In Eq.~(\ref{3fold}) an integration over the directions of the momentum
 ${\bf k}$ is performed.
 To compare with the COSY data \cite{vikomarov}
 one has to integrate the cross section in Eq.~(\ref{3fold})
 over $E_{pp}$ from 0 to 3 MeV and average over
 the neutron cms angle in the interval $\theta_n^*=172^\circ-180^\circ$.
 For the ONE mechanism the square of the spin-averaged
 matrix element takes the form 
\begin{equation}
\label{ONEmatr2}
{\overline {|M_{fi}^{ONE}|^2}}=
\frac{{E_d(E_p+E_n)\varepsilon_p(q)}}{E_p^2}\frac{m^4}{\pi}
 { N}_{pp}^2 \left[u^2(q)+w^2(q)\right ]\,|t(q',k)|^2.
\end{equation}
Eq.~(\ref{ONEmatr2}) is derived on the basis of the relativistic Hamiltonian
dynamics for the three-body problem \cite{bkt}.
The Lorentz-invariant relative momenta
at the vertices $d\to p+n$ and $p+p\to pp (^1S_0)$ are denoted as $q$ and $q'$, 
respectively. The other notations in Eq. (\ref{ONEmatr2}) are:
$E_j$ is the energy of the deuteron ($j=d$), intermediate proton ($p$)
and neutron ($n$) in the cms of the $p+d$ system, and 
$\varepsilon_p(q)=\sqrt{m^2+q^2}$;
$u(q)$ and $w(q)$ are the S- and D-wave components of
the deuteron wave function in momentum space, normalized as
$\int_0^\infty\left (u^2(q)+w^2(q)\right )q^2{dq}={(2\pi)^3}.$
The combinatorial factor $N_{pp}=2 $ in Eq.~(\ref{ONEmatr2}) and also 
the factor $\frac{1}{2}$ in Eq.~(\ref{3fold}) take into account the identity 
of the final protons.
The half-off-shell 
$t$-matrix in the
$^1S_0$ $pp$ state is given by 
\begin{equation}
\label{half}
t(q',k)={-4\pi \int_0^\infty \frac{F_0(q'r)}{q'r}V_{NN}(r)
{\psi^{(-)}}^*(r)r^2dr},
\end{equation}
where $F_0(z)$ is the regular Coulomb function 
for zero orbital momentum  $l=0$ and 
$\psi^{(-)}$
is the scattering wave function obtained from the solution of the
Schr\"odinger equation for a $NN$ potential ($V_{NN}$) 
including the Coulomb interaction ($V_C$),
i.e. $V(r) = V_{NN}(r) + V_C(r)$,  
and normalized as 
$\psi^{(-)}(r)\to \frac{\cos{\delta}}{kr}
\left [F_0(kr)+\tan{\delta} G_0(kr)\right]$.
 Here  $\delta$ is the Coulomb distorted nuclear phase shift and
 $G_l(kr)$ is the irregular
 Coulomb function.
 The pure Coulomb half-off-shell
 t-matrix, $t^c({\bf q},{\bf k})$, derived in Ref. \cite{dolmux},
 gives a very small contribution for $|{\bf k}| \ll|{\bf q}^\prime|$
 and is neglected in the present calculation.
 Further 
 details of the formalism, specifically for evaluating 
 the $\Delta$ and SS mechanisms of
 the breakup reaction $pd\to (pp)n$, can be found in
 Refs.~\cite{uz98,imuz90,uz97}.

\begin{figure}[hbt]
%\mbox{\epsfig{figure=prcd7.ps,height=0.6\textheight, clip=}}
%\mbox{\epsfig{figure=prcdwbacn6.ps,height=0.6\textheight, clip=}}
%\mbox{\epsfig{figure=jh21uz.ps,height=0.8\textheight, clip=}}
\mbox{\epsfig{figure=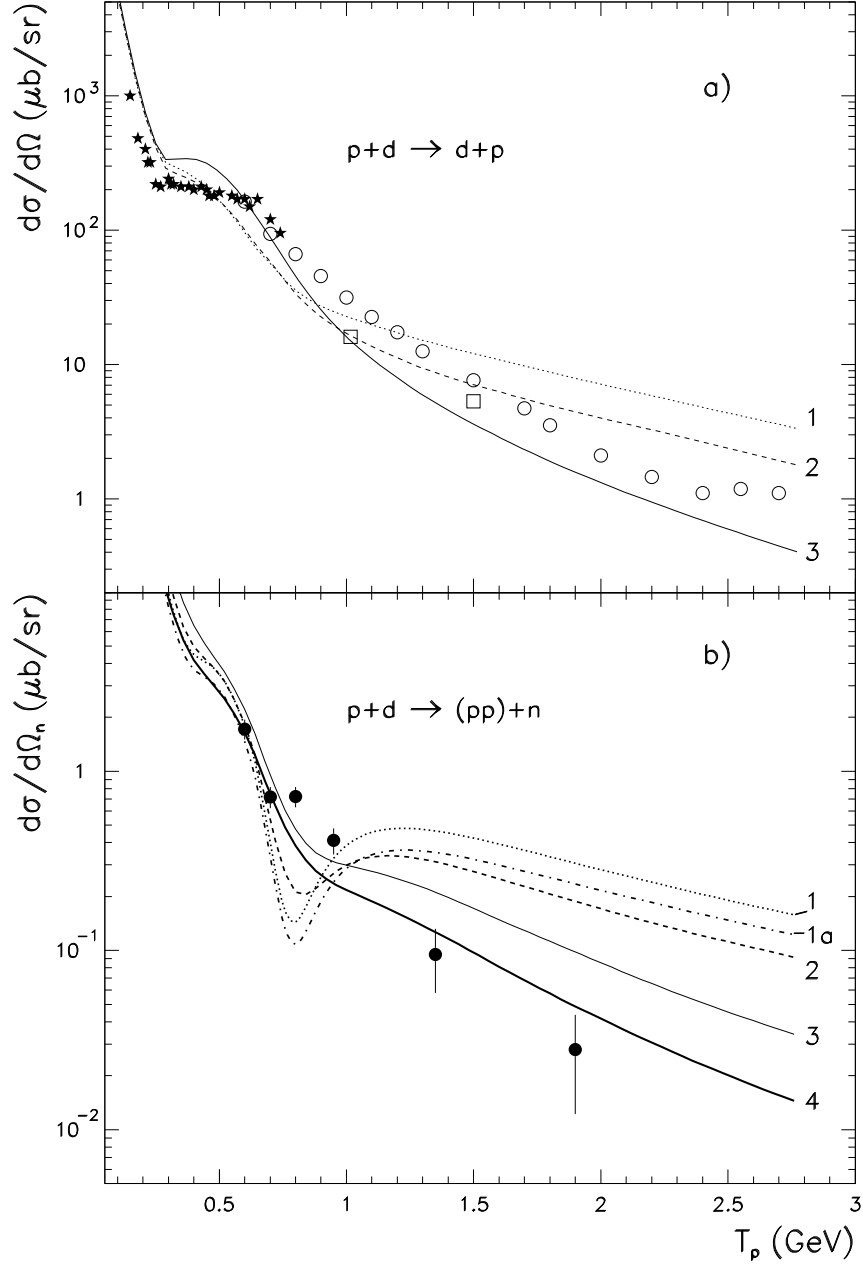,height=0.8\textheight, clip=}}
\caption{
Cms differential
cross sections of the reaction $pd \to dp$ at the proton
scattering angle $\theta^*_p=180^\circ$ (a)
and of the reaction $p+d\to (pp)+ n$
for neutron scattering angles $\theta^*_n=172-180^o$ 
 and relative energies $E_{pp}=0-3$ MeV of the two forward protons 
(b) as a function of the beam energy $T_p$. 
Calculations are performed on the basis of the
ONE+SS+$\Delta$ model described in the text without (lines 1--3)
and with distortions included (4) for the
RSC (1), Paris (2) and the CD Bonn (3,4) potentials.  
The curve 1a shows the result for the RSC potential with
Coulomb effects included. 
Curve 4 is the result for the CD Bonn potential taking into account 
distortions for the ONE mechanism and also Coulomb effects (by the 
suppression factor of 20\%, cf. text).
 Data for $pd\to (pp)n$ are from Ref. \protect\cite{vikomarov},
 and for $pd\to dp$ from \protect\cite{dubal,boud2,berth}.
 }
\label{fig2}
\end{figure}

We start with the reaction $pd\to dp$. Corresponding 
results are shown in Fig.~\ref{fig2}a.
As can be seen, at $T_p<0.5$~GeV  the theoretical predictions
describe the shape of the experimental data but overestimate
the absolute value.
This shortcoming of the model calculation is presumably caused by 
the neglect of the initial and final state interaction. 
Indeed the latter effects were used as an argument in 
Ref. \cite{kls} to introduce a phenomenological distortion factor
of 0.8 for the ONE amplitude to match the 
experimental data on the $pd\to dp$ cross section at
$T_p<0.3$~GeV.
% CHECK
Actually, the distortion effects are energy dependent, therefore
we have refrained from applying such a phenomenological distortion
factor to our results for $pd\to dp$ shown in Fig.~\ref{fig2}a. 
At energies above 1~GeV the results are contradictory. 
While the calculations based on the RSC
and Paris deuteron wave functions lead to an overestimation 
of the cross section we observe an underestimation for the CD
Bonn model. 
In view of expected substantial contributions of heavy nucleon isobars,
which are theoretically not well 
under control at these energies \cite{uz98}, it is not possible to 
draw more concrete conclusions from those deviations at the
present stage. 
Therefore it is much more interesting to look at the
reaction $pd\to (pp)n$ because, as we mentioned above, contributions 
from such isobar states are expected to be suppressed in the breakup 
channel and, consequently, a comparison
of a model calculation with the breakup data should provide a 
much more conclusive test for the $NN$ interaction models.

\begin{figure}[hbt]
\mbox{\epsfig{figure=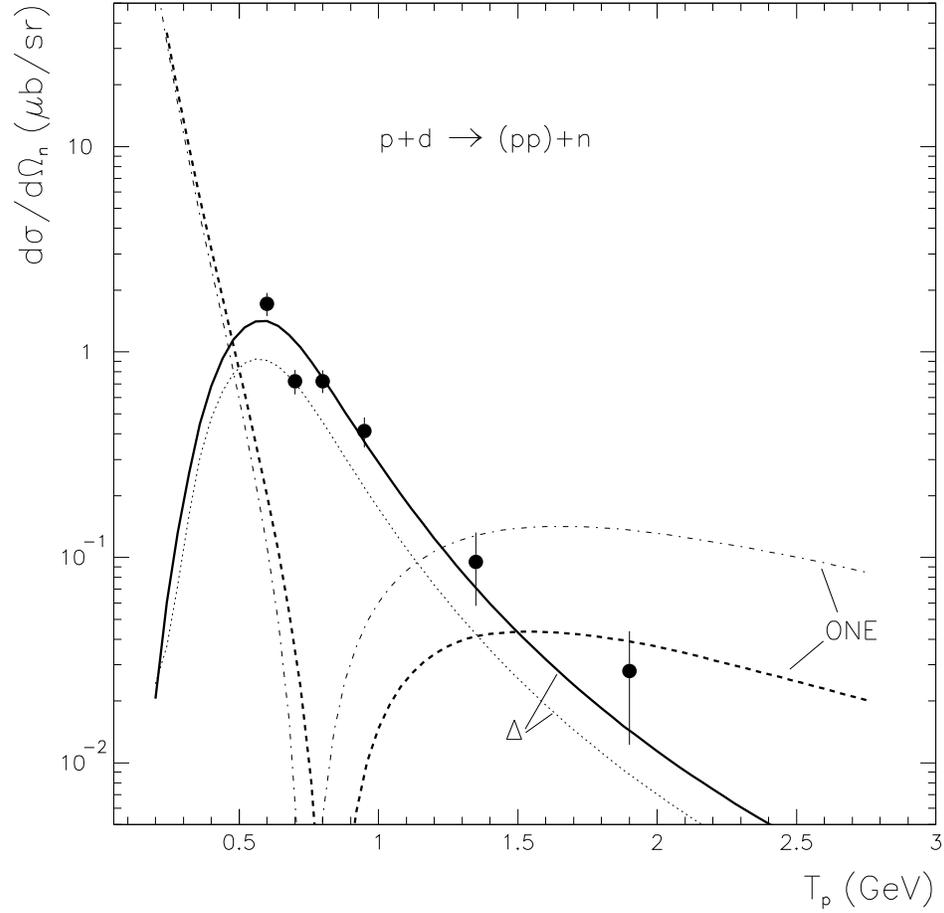,height=0.6\textheight, clip=}}
\caption{Contributions of the considered reaction
 mechanisms to the 
 cms differential cross section of the
 reaction $p+d\to (pp)+ n$
 at neutron scattering angle $\theta^*_n=172-180^o$
 and relative energies $E_{pp}=0-3$ MeV of the two forward protons.
Results are shown for the Paris 
(ONE -- dash-dotted line, $\Delta$ -- dotted line)
and CD Bonn
(ONE -- dashed line, $\Delta$ -- solid line)
$NN$ potentials.
Data are from Ref. \protect\cite{vikomarov}.
 }
\label{fig3}
\end{figure}

Our predictions for the breakup reaction $pd\to (pp)n$ are
shown in Fig.~\ref{fig2}b.  Let us focus first on the results
without Coulomb interaction and without distortion effects.
 One can see from this figure, the dip becomes  less pronounced in
 the ONE+SS+$\Delta$ model calculation when one comes from the 
 RSC (line 1) to Paris (line 2) and then to CD Bonn (line 3) $NN$ 
 potential. At the same time the tail of the cross
 section at higher energies $T_p>1 $ GeV becomes smaller. 
 As a whole, the model predictions  with the CD Bonn wave functions 
 are much closer to the data  than for the RSC and Paris potentials.
%
%It is evident that the ONE+SS+$\Delta$ 
%model calculation employing the CD Bonn wave functions 
%(line 3) yields results that are much closer to the data 
%than the calculations based on the RSC (line 1) and Paris $NN$ 
%potentials (line 2). 
%
 The improvement is caused primarily by the relative softness of the CD Bonn 
 wave functions in the $^3S_1-^3D_1$ and $^1S_0$ states
 as compared to those of the RSC and Paris potentials.
 Because of this feature the relative importance of all
 mechanisms in question are significantly changed when using the CD Bonn 
 interaction instead of the Paris or RSC models.
 This is demonstrated more explicitly in 
 Fig. \ref{fig3}, where the separate contributions of the 
 ONE+SS+$\Delta$ model are shown
 for the CD Bonn wave function and also for the Paris potential.
 Obviously the magnitude of the
 ONE cross section at $T_p>0.8$ GeV for the CD Bonn potential is 
 considerably smaller (by factor of 3-4)
 as compared to the result with the Paris potential (cf.  Fig. \ref{fig3}).
 This reduction of the ONE contribution at higher energies
 leads to a much better agreement between the ONE+SS+$\Delta$ model
 and the data at $T_p> 1$ GeV for this $NN$ model (see Fig. \ref{fig2}b). 
 At the same time the $\Delta$ contribution is larger for the CD Bonn
 interaction as compared to the Paris model. As mentioned above, 
 the CD Bonn wave functions decrease much faster with increasing momentum
 $q$ as compared to those of the Paris and RSC potentials. This means
 that, in configuration space, the CD Bonn
 wave functions provide a higher probability density 
 $|\psi(r)|^2$ for finding the $NN$ system at short distances  
 ($r<0.5$ fm) as compared to the case of the Paris (or RSC) potentials.
 Since the amplitude of the $\Delta$ mechanism
 is proportional to the averaged value of $r^{-2}$ \cite{kls,imuz90},
 it is clear that the $\Delta$ contribution will be larger 
 for the CD Bonn interaction model than for the Paris potential. 
 This property of the $\Delta$ mechanism was demonstrated
 earlier in Ref. \cite{umuzsc89} by comparing its contribution
 to the $pd\to dp$ cross section calculated 
 with the RSC and Paris potentials, respectively.
 The increase of the $\Delta$ contribution to the
 breakup cross section by approximately 50\%  for the
 CD Bonn potential fills in much of the 
 ONE dip \cite{imuz90,smuz98,uz2002} of the cross section (cf. 
Fig. \ref{fig2}b) and, as a result, improves the agreement with data.
 One can also see from Fig.~\ref{fig3}
 that for the CD Bonn interaction the $\Delta$ mechanism alone 
 describes the measured breakup cross section already rather well
 at $T_p>0.8$~GeV, but the interferences
 between the $\Delta$ and the SS mechanisms at around 0.8 GeV and 
 between the $\Delta$ and ONE amplitudes for $T_p>1$~GeV
 destroy this agreement (see Fig. \ref{fig2}b).
 The SS contribution (not shown in Fig. \ref{fig3}) is relatively small itself.
 But its contribution to the breakup cross section is also smaller for CD Bonn than 
 for the Paris potential. 
 
% CHECK 
For a direct comparision with the experiment let us now also take into
account effects of the initial and final state interaction in the ONE contribution.
This is done in distorted wave Born approximation (DWBA)
as described in Refs. \cite{uz2002,uz97}. The corresponding result 
for the CD Bonn model is shown by the thick solid line in Fig.~\ref{fig2}b. 
In addition, in this calculation the Coulomb effects are included.
Since the CD Bonn potential is given only in momentum space it is rather
difficult to account for the Coulomb interaction rigorously \cite{Elster}.
However, by performing corresponding calculations for the 
RSC and Paris potentials in configuration space 
we found that inclusion of the Coulomb interaction in the final $pp$
system decreases the ONE+SS+$\Delta$ cross section  
by approximately 20\% (cf. the curves 1 and 1a in Fig.~\ref{fig2}b
for the RSC potential). 
Thus, we have assumed here that the Coulomb repulsion produces the same 
suppression of the cross section also for the CD Bonn $NN$ interaction.
Obviously, including both these effects leads
to a further reduction of the predicted cross section. In particular,
now the total result for the ONE+SS+$\Delta$ model is already in 
qualitative agreement with the breakup data.

\begin{figure}[hbt]
\mbox{\epsfig{figure=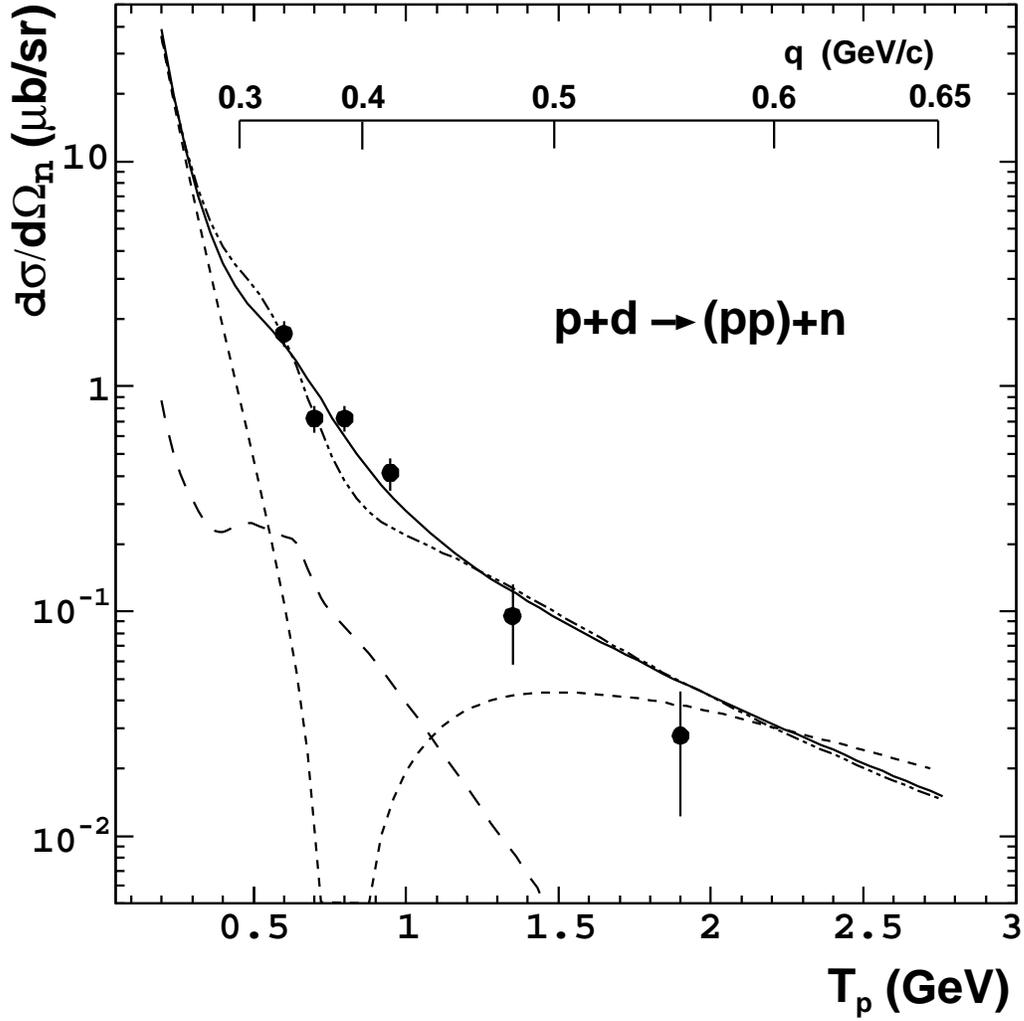,height=0.6\textheight, clip=}}
\caption{The same as in Fig. \protect\ref{fig3}, but 
 using the CD Bonn $NN$ potential only.
ONE -- short-dashed line; SS -- long-dashed line;
coherent sum of ONE+$\Delta$ with Coulomb effects included -- solid line;
ONE+SS+$\Delta$ with Coulomb effects included --- dashed-double dotted line.
In the latter two cases distortions in the ONE contribution
are also included. The upper scale shows the internal momentum of the
nucleons in the deuteron for the ONE, cf. Eq. (\ref{ONEmatr2}).}
\label{fig4}
\end{figure}

 Let us make some further comments.
 To begin with, the contribution of the SS mechanism is presumably
 overestimated. This contribution involves the subprocess of 
 $pn\to pn$ scattering which takes place completely off-shell.
 However, in the actual calculation the on-shell
 $pn\to pn$ amplitude is used \cite{kls,imuz90}.
 In addition, double elastic $pN$ scattering, which is not
 considered in this paper, should also reduce the influence
 of the SS mechanism. Actually,
 as was shown in Ref. \cite{gurvitz80}, the coherent sum of
  single and double $pN$ scattering produces
 a $pd\to dp$ cross section which is considerably smaller than
 that for the SS alone.
 None the less, 
 we should stress in this context that backward elastic $pd$ 
 scattering is not very sensitive to the details of the SS mechanism.
 At $T_p< 0.8$ GeV, this
 reaction is dominated by the ONE and $\Delta$ mechanisms. 
 Indeed, omitting the SS contribution from the
 ONE+SS+$\Delta$ model practically does not change the $pd\to dp$ 
 cross section \cite{bdillig}. 
 The situation is completely different for the breakup reaction 
 $pd\to (pp)n$, specifically around the dip structure because
 there the contribution of the ONE amplitude is basically zero. 
 Here the ONE+$\Delta$ result 
 differs significantly from the one based on the ONE+SS+$\Delta$ 
 amplitude in the region of 0.6-0.9 GeV and as a matter of facts 
 describes the data better in this region (see Fig. \ref{fig4}).
 Possible 
 non-nucleonic contributions, like the $\Delta \Delta$ and $NN^*$ components
 of the deuteron and diproton, can also change the results in
 the region of the expected dip.
  The role of the relativistic $P$-wave component that couples to the 
  $^1S_0$ state was studied recently in Ref.~\cite{kaptari}   
in a covariant Bethe-Salpeter approach. 
  According to Ref.~\cite{kaptari} the contribution of this $P$-wave 
 completely masks the dip of the $pd\to (pp)n$ cross section 
 and makes the $pp$ scattering amplitude properly small
 to achieve agreement with the experiment at higher energies.
 However, only the ONE mechanism was discussed in Ref. \cite{kaptari} and, 
 moreover, without rescattering effects. The inclusion of the $\Delta$ 
 contribution, 
 may change the result obtained in Ref. \cite{kaptari} significantly.
 As was shown here and in Ref. \cite{ukrs02}, already the $\Delta$ mechanism
 alone is sufficient to describe the data.

 In conclusion, we analyzed the deuteron breakup data $pd\to (pp)n$ in the 
 framework of the ONE+SS+$\Delta$ model that has been previously applied 
 to backward elastic $pd$ scattering in a similar kinematical region. 
 We show that the unpolarized cross section of this reaction is very sensitive
 to the behaviour of the $NN$ interaction at short distances as reflected
 in the high momentum components of the deuteron and $pp$ wave functions. 
Due to the relative 
smallness of the high momentum component of the CD Bonn 
 wave functions in the $^3S_1-^3D_1$ and $^1S_0$ states 
 a much better agreement with the breakup data is achieved than for 
 models with harder wave functions like the RSC or Paris potentials. 
 Future polarization measurements in the reaction $pd\to (pp)n$ can
 provide a further tests for the present picture of the $pd$ interaction
 and the $NN$ dynamics at short distances.

{\bf Acknowledgments.} One of us (Yu.N.U.) gratefully acknowledges 
 the warm hospitality
 at the IKP-2 of the Forschungszentrum J\"ulich, where part of this work was done.
 This work was supported by the BMBF project
 grant KAZ 99/001 and the Heisenberg-Landau program.

\eject
%%%%%%%%%%%%%%%%%%%%%%%
\eject 
%%%%%%%%%%%%%%%%%%%%%%%%%%%%%%%%%%%%%%%%%%%%%%%%%%%%%%%%%%%%%
%%%%%%%%%%%%%%%%%%%%%%%
\eject 
%%%%%%%%%%%%%%%%%%%%%%%%%%%%%%%%%%%%%%%%%%%%%%%%%%%%%%%%%%%%%
%%%%%%%%%%%%%%%%%%%%%%%
\eject 
%%%%%%%%%%%%%%%%%%%%%%%%%%%%%%%%%%%%%%%%%%%%%%%%%%%%%%%%%%%%%
\end{document}